\documentclass[fleqn,12pt,twoside]{article}
\usepackage[headings]{espcrc1}

\usepackage{graphicx}

\title{Correlations in the Parton Recombination Model}

\author{S.A. Bass\address[Duke]{Department of Physics, Duke University,
                Durham, NC 27708-0305, USA}
        \address[RIKEN]{RIKEN BNL Research Center,
        Brookhaven Nat. Lab., Upton, NY 11973, USA},
	R.J. Fries \address[UMN]{School of Physics and Astronomy, 
	Univ. of Minnesota, Minneapolis, MN 55455, USA}
        and B. M\"uller\addressmark[Duke]}

\runtitle{Correlations in the Parton Recombination Model}
\runauthor{S.A. Bass, R.J. Fries and B. M\"uller}

\begin{document}

\maketitle

\begin{abstract}
We describe how parton recombination can address the recent measurement
of dynamical jet-like two particle correlations. In addition we
discuss  the possible effect realistic
light-cone wave-functions including higher Fock-states may
have on the well-known elliptic flow valence-quark
number scaling law.
\end{abstract}

\section{Introduction}
Recent data from the Relativistic Heavy Ion Collider (RHIC) have shown a
strong nuclear suppression of the pion yield at transverse momenta larger than
2 GeV/$c$ in central Au + Au collisions, compared to $p+p$ interactions
\cite{PHENIX}. This is widely seen as the experimental confirmation of
jet quenching, the phenomenon that high energy partons lose energy when
they travel through the hot medium created in a heavy ion collision
\cite{GyulWang:94,BDMS:01,Muller:02}, entailing a suppression of intermediate
and high $P_T$ hadrons.
                                                                                
However, the experiments at RHIC have provided new puzzles. The amount of
suppression seems to depend on the hadron species. In fact, in the production
of protons and antiprotons between 2 and 4 GeV/$c$ the suppression seems to
be completely absent. Generally, pions and kaons appear to suffer from a
strong energy loss while baryons and antibaryons do not. Two stunning
experimental facts exemplify this \cite{PHENIX-B,STAR-B,STAR-L,PHENIX-L}.
First, the ratio of protons over positively charged pions is equal or above
one for $P_T > 1.5 {\rm ~GeV}/c$ and is approximately constant up to 4 GeV/$c$.
Second, the nuclear suppression factor $R_{AA}$ below 4 GeV/$c$ is close to
one for baryons, while it is about 0.3 for mesons.
                                                                                
There have been recent attempts to describe the different behavior of baryons
and mesons through the existence of gluon junctions \cite{Kharzeev:1996sq,GyulVit:01}
or alternatively through
recombination as the dominant mechanism of hadronization
\cite{Fries:2003vb,Greco:2003xt,Hwa:2002tu}.
The recombination picture has attracted additional attention due to the
observation that the elliptic flow pattern of different hadron species can
be explained by a simple recombination mechanism
\cite{Molnar:2003ff,Fries:2003kq,Greco:2003mm}.
The anisotropies $v_2$ for the different hadrons
in the $p_t/n$ range ($n$ being the number of valence quarks of the hadron) 
of 1.0 - 2.5~GeV are compatible with a
universal value of $v_2$ in the parton phase, related to the hadronic flow
by factors of two and three depending on the number of valence quarks
\cite{Sorensen:03}.
                                                                                
The competition between recombination and fragmentation
delays the onset of the perturbative/fragmentation regime to relatively high
transverse momentum of 4--6 GeV/$c$, depending on the hadron species, providing
a natural explanation for the aforementioned phenomena. To this date,
parton recombination has developed into the most successful model
for describing hadron production at RHIC in the intermediate
transverse momentum domain.

\section{Two-Particle Correlations}

One of the biggest challenges for the recombination models to date
has been the measurement of dynamic two-particle correlations.
The picture of quarks recombining from a collectively flowing,
deconfined thermal quark plasma appears to be at odds with the
observation of ``jet-like'' correlations of hadrons observed
in the same transverse momentum range of 2 to 5 GeV/c
\cite{STAR:03corr,Sick:04}.
The experiments at RHIC measure the associated particle yield per trigger
hadron $A$. After subtracting the uncorrelated background and using the
notation $\Delta\phi = |\phi_A - \phi_B|$, the relevant observable is
defined as
\begin{equation}
  Y_{AB}(\Delta \phi) = N_A^{-1} \left(
  {\frac{dN_{AB}}{d(\Delta \phi)} - \frac {d (N_A N_B)}{d(\Delta \phi)}}
\right).
\end{equation}

Triggering on a hadron, e.g., with transverse momentum
2.5 GeV/$c < p_T <$ 4 GeV/$c$, the data shows an enhancement
of hadron emission in a narrow angular cone around the
direction of the trigger hadron in a momentum window below
2.5 GeV/$c$. Can such correlations be reconciled with the claim
that hadrons in this momentum range are mostly created by
recombination of quarks?
                                                                                
\begin{figure}[htb]
\begin{minipage}[t]{9cm}
\includegraphics[width=9cm]{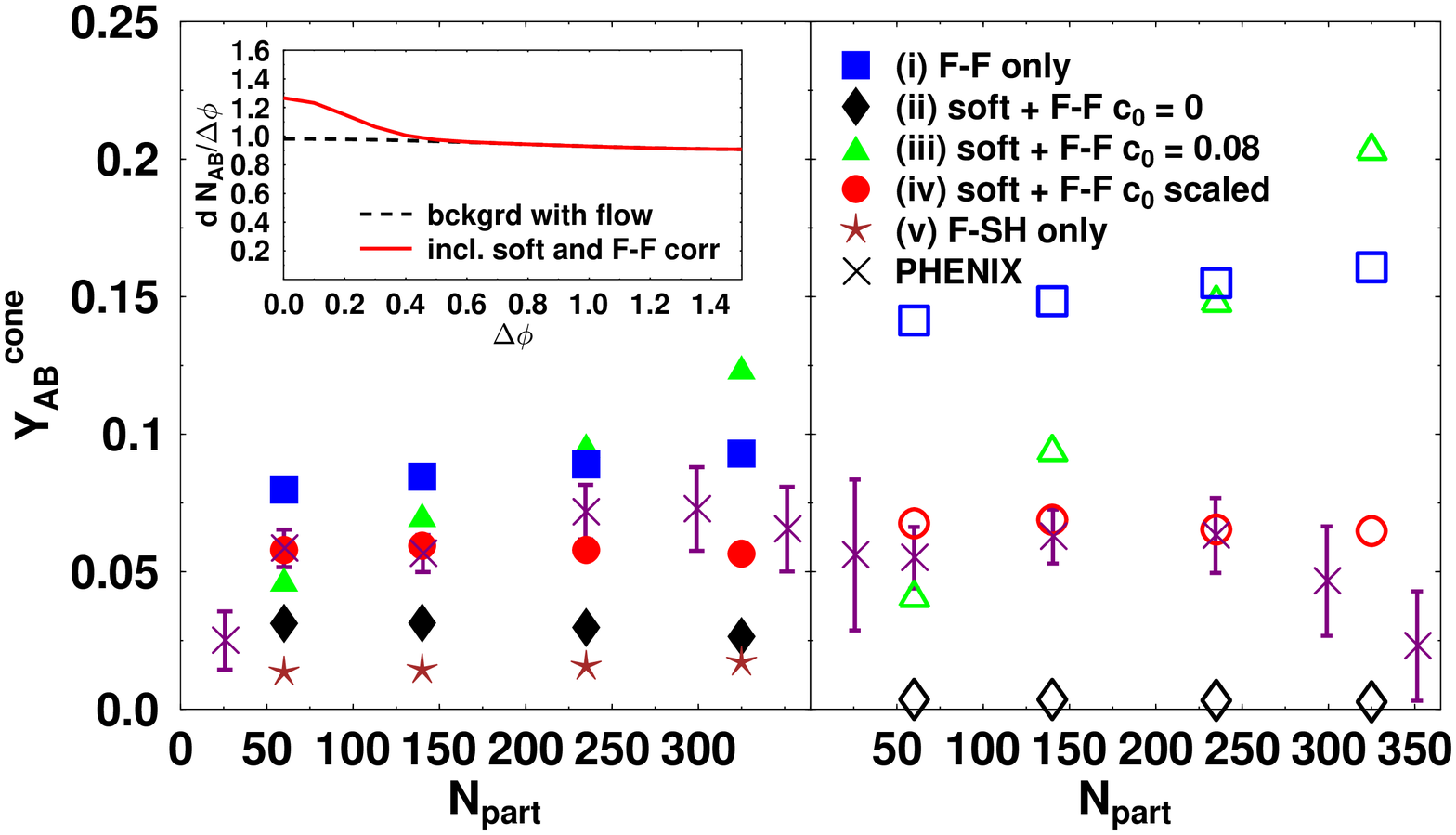}
\caption{$Y_{AB}^{\rm cone}$ which is $Y_{AB}$
  integrated over $0\le \Delta\phi \le 0.94$, for meson (left panel) and
  baryon triggers (right panel) as a function of centrality. 
  The inset shows the
  associated yield as a function of $\Delta \phi$ at an impact parameter $b=8$ fm.}
\label{fig1}
\end{minipage}
\hspace{\fill}
\begin{minipage}[t]{6.5cm}
\includegraphics[width=6.2cm]{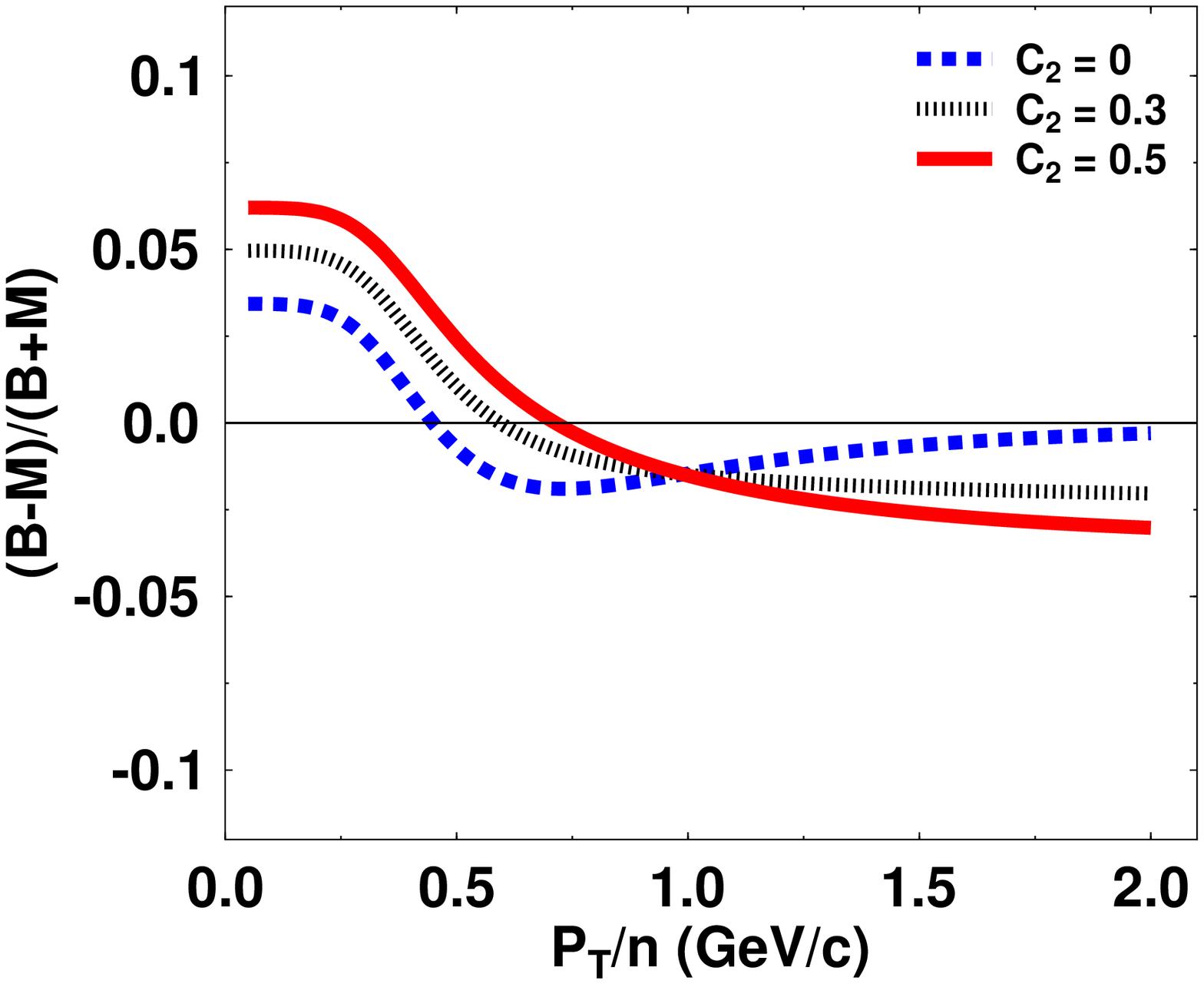}
\caption{Relative difference $(\tilde v_2^{\rm (B)}
-\tilde v_2^{\rm (M)})/(\tilde v_2^{\rm (B)}+\tilde v_2^{\rm (M)})$
between the scaled meson and baryon elliptic flow for three different
sizes of the higher Fock state component.}
\label{fig2}
\end{minipage}
\end{figure}

Obviously, the existence of such correlations 
is incompatible with any model assuming that no correlations 
exist among the quarks before recombination, since
such correlations require deviations
from a global thermal equilibrium in the quark phase.
However,  it can be shown that correlations among partons in a
quark-gluon plasma naturally translate into correlations
between hadrons formed by recombination of quarks \cite{Fries:2004hd}.
Correlations are even enhanced by an amplification
factor $Q=n_A n_B$ similar to the scaling of elliptic flow.
The interaction of hard partons with the medium has been
discussed as one plausible mechanism for the existence of
such parton correlations, even though other scenarios for
the creation of parton-parton correlations in the deconfined
phase are possible. A numerical example displayed in figure~\ref{fig1} shows
that two-parton correlations of order $\approx 10\%$ will be
sufficient to explain hadron correlations as measured by the
PHENIX collaboration. One may conclude that the existence of
localized angular correlations among hadrons are not
in contradiction with the recombination scenario but rather 
indicative for the existence of correlations among 
quarks prior to hadronization.

\section{Beyond the Valence Quark Approximation}
Recombination models usually are based on the concept of {\em constituent
quark recombination}, which assumes that the probability for the emission
of a hadron from a deconfined medium is proportional to the probability
for finding the valence quarks of the hadron in the density matrix
describing the source. The baryon enhancement, as well as the different
momentum dependence of meson and baryon anisotropies, rely essentially
on the different number of valence quarks in mesons (two) and baryons
(three). The simplicity of this concept has been criticized, because
it does not do justice to the complexity of the internal structure of
hadrons in quantum chromodynamics (QCD). The question is how a more
realistic  treatment of  the internal structure of hadrons
affects these observables.

In the light-cone frame, where formally the
hadron momentum $P \to \infty$ and the momentum fractions of the partons
are the only dynamic degrees of freedom, a meson $M$ with valence quarks
$q_\alpha$ and $\bar q_\beta$ can then be written as an expansion in
terms of increasingly complex Fock states:

\begin{eqnarray}
|M\rangle
&=& \int_0^1 dx_a dx_b \delta(x_a+x_b-1)
          c_1(x_a,x_b) \left|q_\alpha(x_a){\bar q}_\beta(x_b)\right\rangle
\nonumber \\
&+& \int_0^1 dx_a dx_b dx_c \delta(x_a+x_b+x_c-1)
          c_2(x_a,x_b,x_c) \left|q_\alpha(x_a){\bar q}_\beta(x_b)
          g(x_c)\right\rangle
\\
&+& \int_0^1 \prod\limits_{i=a}^d dx_i \,\, \delta\left(\sum\limits_{i=a}^d x_i-1\right)
          c_3(x_a,x_b,x_c,x_d)
          \left|q_\alpha(x_a){\bar q}_\beta(x_b)q_\gamma(x_c)
                 {\bar q}_\gamma(x_d)\right\rangle + \ldots
\nonumber
\label{eq01}
\end{eqnarray}

It has been shown \cite{Muller:2005pv} that the yield of relativistic 
parton clusters is independent of the number of partons in the cluster. 
Therefore, hadron spectra remain
unaffected by the inclusion of a higher Fock state with an additional gluon. 
One important
implication is that gluon degrees of freedom could be accommodated
during hadronization. They simply become part of the
quark-gluon wave functions of the produced hadrons, but remain
hidden constituents because the produced hadrons do not
contain valence gluons.
                                                                                     
However, higher Fock states introduce deviations from the
scaling law for elliptic flow. 
Using a narrow wave function limit, one can easily generalize 
the well-known valence quark scaling law 
to higher Fock states:
\begin{equation}
  v_2^{\rm (H)}(P) \approx \sum_\nu | c_\nu |^2 n_\nu v_2(P/n_\nu)
\label{eq08}
\end{equation}
Figure \ref{fig2} shows the relative difference $(\tilde v_2^{\rm (B)}
-\tilde v_2^{\rm (M)})/(\tilde v_2^{\rm (B)}+\tilde v_2^{\rm (M)})$
between the scaled meson and baryon elliptic flow for three different
sizes of the higher Fock state component (0\%, 30\%, 50\%). In all cases,
baryons have a slightly larger scaled $\tilde v_2$ than mesons at small
momenta. This effect is likely to be overwhelmed by the influence of mass
differences, which have been neglected in the sudden recombination model.
At larger momenta, the scaled meson $\tilde v_2$ is slightly larger.
In principle, these violations on the order of $\sim 10\%$ should be visible in
a scaling analysis and first observations along these lines have
been reported at this meeting \cite{soerensen_qm05}.

It should be emphasized that the interpretation of elliptic flow data from RHIC
proving the existence of quark degrees of freedom in the bulk matter
is still valid. However, the
connection of the measured elliptic flow to the quark elliptic flow
might be less straight forward than anticipated.

This work was supported in part by RIKEN, the Brookhaven National
Laboratory, and DOE grants DE-FG02-05ER41367, DE-AC02-98CH10886
and DE-FG02-87ER40328.
S.A.B. acknowledges support from a DOE Outstanding Junior Investigator
Award.


\begin{thebibliography}{9}

\bibitem{PHENIX}
K.~Adcox {\it et al.}  [PHENIX Collaboration],
Phys.\ Rev.\ Lett.\  {\bf 88} (2002), 022301;
C.~Adler {\it et al.}  [STAR Collaboration],
Phys.\ Rev.\ Lett.\  {\bf 90} (2003), 082302.
                                                                                
\bibitem{GyulWang:94}
J.~D.~Bjorken,
FERMILAB-PUB-82-059-THY (1982);
M.~H.~Thoma and M.~Gyulassy,
Nucl.\ Phys.\ B {\bf 351} (1991), 491;
X.~N.~Wang and M.~Gyulassy,
Phys.\ Rev.\ Lett.\  {\bf 68} (1992), 1480;
R.~Baier, Y.~L.~Dokshitzer, A.~H.~Mueller, S.~Peigne and D.~Schiff,
Nucl.\ Phys.\ B {\bf 483} (1997), 291;
M.~Gyulassy, P.~Levai and I.~Vitev,
Phys.\ Rev.\ Lett.\  {\bf 85} (2000), 5535;
U.~A.~Wiedemann,
Nucl.\ Phys.\ B {\bf 588} (2000), 303.
                                                                                

\bibitem{BDMS:01}
R.~Baier, Y.~L.~Dokshitzer, A.~H.~Mueller and D.~Schiff,
JHEP {\bf 0109} (2001), 033;
                                                                                
\bibitem{Muller:02}
B.~Muller,
Phys.\ Rev.\ C {\bf 67} (2003), 061901.
                                                                                
\bibitem{PHENIX-B}
K.~Adcox {\it et al.}  [PHENIX Collaboration],
Phys.\ Rev.\ Lett.\  {\bf 88} (2002), 242301.
                                                                                
\bibitem{STAR-B}
C.~Adler {\it et al.}  [STAR Collaboration],
Phys.\ Rev.\ Lett.\  {\bf 86} (2001), 4778.
                                                                                
\bibitem{STAR-L}
C.~Adler {\it et al.}  [STAR Collaboration],
Phys.\ Rev.\ Lett.\  {\bf 89} (2002), 092301.
                                                                                
\bibitem{PHENIX-L}
K.~Adcox {\it et al.}  [PHENIX Collaboration],
Phys.\ Rev.\ Lett.\  {\bf 89} (2002), 092302.

\bibitem{Kharzeev:1996sq}
  D.~Kharzeev,
  Phys.\ Lett.\ B {\bf 378}, 238 (1996).
                                                                                
\bibitem{GyulVit:01}
I.~Vitev and M.~Gyulassy,
Phys.\ Rev.\ C {\bf 65} (2002), 041902.
                                                                                


\bibitem{Fries:2003vb}
  R.~J.~Fries, B.~Muller, C.~Nonaka and S.~A.~Bass,
  Phys.\ Rev.\ Lett.\  {\bf 90}, 202303 (2003).

\bibitem{Greco:2003xt}
  V.~Greco, C.~M.~Ko and P.~Levai,
  Phys.\ Rev.\ Lett.\  {\bf 90}, 202302 (2003).


\bibitem{Hwa:2002tu}
  R.~C.~Hwa and C.~B.~Yang,
  Phys.\ Rev.\ C {\bf 67}, 034902 (2003).

\bibitem{Molnar:2003ff}
  D.~Molnar and S.~A.~Voloshin,
  Phys.\ Rev.\ Lett.\  {\bf 91}, 092301 (2003).


\bibitem{Fries:2003kq}
  R.~J.~Fries, B.~Muller, C.~Nonaka and S.~A.~Bass,
  Phys.\ Rev.\ C {\bf 68}, 044902 (2003).

\bibitem{Greco:2003mm}
  V.~Greco, C.~M.~Ko and P.~Levai,
  Phys.\ Rev.\ C {\bf 68}, 034904 (2003).

\bibitem{Sorensen:03}
P.~Sorensen for the STAR Collaboration,
J.\ Phys. {\bf G30} (2004) S217.


\bibitem{STAR:03corr}
C.~Adler {\it et al.}  [STAR Collaboration],
Phys.\ Rev.\ Lett.\  {\bf 90}, 082302 (2003);
                                                                                
\bibitem{Sick:04}
A.~Sickles for the PHENIX Collaboration,
J.\ Phys. {\bf G30} (2004) S1291.
 
\bibitem{Fries:2004hd}
  R.~J.~Fries, S.~A.~Bass and B.~Muller,
  Phys.\ Rev.\ Lett.\  {\bf 94}, 122301 (2005).


\bibitem{Muller:2005pv}
  B.~Muller, R.~J.~Fries and S.~A.~Bass,
  Phys. Lett. B {\bf 618}, 77 (2005).

\bibitem{soerensen_qm05}
P.~Sorensen, these proceedings.

\end{thebibliography}
\end{document}